\documentclass[nofootinbib,prd,twocolumn,showpacs,showkeys,preprintnumbers]{revtex4}
\usepackage{hyperref,amssymb,amsmath,mathrsfs,bm,graphicx}
\begin{document}
\title {Dissipative collapse of axially symmetric, general relativistic, sources: A general framework and some applications}
\author{L. Herrera}
\email{lherrera@usal.es}
\affiliation{Escuela de F\'\i sica, Facultad de  Ciencias,
Universidad Central de Venezuela, Caracas 1050, Venezuela}
\author{A. Di Prisco}
\email{adiprisc@fisica.ciens.ucv.ve}
\affiliation{Escuela de F\'\i sica, Facultad de  Ciencias,
Universidad Central de Venezuela, Caracas 1050, Venezuela}
\author{J. Ib\'a\~nez  }
\email{j.ibanez@ehu.es}
\affiliation{Departamento de F\'\i sica Te\'orica e Historia de la Ciencia,
Universidad del Pa\'{\i}s Vasco, Bilbao 48993, Spain}
\author{J. Ospino}
\email{jhozcrae@usal.es}
\affiliation{Departamento de Matem\'atica Aplicada,  Universidad de Salamanca, Salamanca 37007, Spain.}
\date{\today}
\begin{abstract}
We carry out a general  study on the collapse of axially (and reflection) symmetric sources in the context of general relativity. All  basic equations and concepts required to perform such a general study  are deployed. These  equations  are written down for a general anisotropic dissipative fluid. The proposed approach allows for  analytical studies as well as for numerical applications. A causal transport equation derived from the Israel-Stewart theory is applied, to discuss some thermodynamic aspects of the problem. A set of scalar functions (the structure scalars) derived from the orthogonal splitting of the Riemann tensor  are calculated and their role in the dynamics of the source is clearly exhibited. The characterization of the gravitational radiation emitted by the source is  discussed.
 \end{abstract}
\date{\today}
\maketitle

\section{Introduction}
In a recent paper we have presented a general framework for studying axially symmetric static sources \cite{1}. The physical arguments supporting the study of axially symmetric sources were clearly exposed there, accordingly we shall not repeat them here. 

We intend in this work to extend the above mentioned study to the fully dynamic case. The reasons to undertake such an endeavour are easy to understand.

 Indeed, the static (and  quasi--static) approximation is very sensible because the hydrostatic time scale is very small for many phases of the life of a star. Thus, it is of the order of 27 minutes for the sun, 4.5 seconds for a white dwarf and $10^{-4}$ seconds for a neutron star of one solar mass and $10$ Km radius \cite{2, 3, 4}. However, during their evolution, self--gravitating objects may pass through phases of intense dynamical activity, with time scales of the order of magnitude of (or even smaller than) the hydrostatic time scale, and  for which the static (quasi--static) approximation is clearly not reliable, e.g., the collapse of very massive stars \cite{5}, and the quick collapse phase preceding neutron star formation, see for example \cite{6} and references therein. In these cases it is mandatory to take into account terms which describe departure from equilibrium, i.e. a full dynamic description has to be used.

Analytical approaches to describe the evolution of  axially (and reflection) symmetric self--gravitating  fluids have been proposed before \cite{7bis, 8bis, 9bis}. However in these latter references only perfect fluids were considered, and furthermore the source was described in Bondi, null, coordinates \cite{7, 8}. However, the perfect fluid condition seems to be a too stringent restriction  for axially symmetric sources, even in the static case \cite{1, prueba}. On the other hand,   Bondi coordinates are known to be very useful for the treatment  of  gravitational radiation in vacuum, but are not particularly suitable  within the source. An analytical approach, which shares some similarities with ours,   although  restricted to the perfect fluid case, may be found in \cite{Z}.

 Therefore, here, we propose a $1+3$ approach, and the source under consideration is as general as possible. Including all non--vanishing stresses compatible with the symmetry of the problem, as well as dissipative phenomena.

The relevance of dissipative processes in the study of gravitational collapse cannot be overemphasized. Indeed, dissipation
 due to the emission of massless
particles (photons and/or neutrinos) is a characteristic process in the
evolution of massive stars. In fact, it seems that the only plausible
mechanism to carry away the bulk of the binding energy of the collapsing
star, leading to a neutron star or black hole, is neutrino emission
\cite{1d}. 

We shall describe dissipation in the diffusion approximation. This
assumption is in general very sensible, since the mean free path of
particles responsible for the propagation of energy in stellar
interiors is in general very small as compared with the typical
length of the object.
Thus, for a main sequence star as the sun, the mean free path of
photons at the centre, is of the order of $2$ cm. Also, the
mean free path of trapped neutrinos in compact cores with densities
about $10^{12}$ g. cm$^{-3}$ becomes smaller than the size of the stellar
core \cite{3d, 4d}.

Furthermore, the observational data collected from supernova 1987A
indicates that the regime of radiation transport prevailing during the
emission process, is closer to the diffusion approximation than to the
streaming out limit \cite{5d}.

On the other hand,  the inclusion of pressure anisotropy  is based on the fact that the local anisotropy of pressure may be caused by a large variety of physical phenomena, of the kind  we expect in compact objects (see Ref. \cite{14a, hdmost04, hmo02, hod08, hsw08, p1, p2, p3} and references therein  for an extensive discussion on this point). 

Among all possible sources of anisotropy,  there are two  particularly related to our primary interest. The first one is the intense magnetic field observed in compact objects such as white dwarfs,  neutron stars, or magnetized strange quark stars  (see, for example, Refs. \cite{15a, 16a, 17a, 18a, 19a} and references therein). Indeed, it is a well-established fact  that  a magnetic field acting on a Fermi gas produces pressure anisotropy (see  Refs. \cite{ 23a, 24a, 25a, 26a} and references therein). In some way, the magnetic field can be addressed as a fluid anisotropy. 

 Another source of anisotropy expected to be present  in neutron stars and, in general, in highly dense matter,  is the viscosity (see \cite{Anderson, sad, alford, blaschke, drago, jones, vandalen, Dong} and references therein).

To carry out the program sketched above, we shall apply the $1+3$ formalism developed in \cite{21cil, n2, refe1, n3, n1, refe2, refe3, 22cil, nin}, (not to confound with the $3+1$ formalism used in numerical general reltivity) coupled to the Israel-Stewart transport equation, within the context of axial symmetry. However, in spite of its advantages (e.g. coordinate independence and completeness \cite{n1}), we shall not follow here a frame formalism but a coordinate basis approach in which the orthonormal frame is only used to identify frame components of proper vectors as scalars that can have a covariant interpretation. The reason for proceeding in this way is not related to any specific advantage of our approach, with respect  to the tetrad formalism, but rather by the simple  fact that having  been working in the past, with the former \cite{9, 14}, we are more familiar  with it.

Besides the great complexity of the equations, the  setup of the presented framework, faces another important challenge, namely: the fact that the source should  emit gravitational radiation.  Indeed, the gravitational collapse even if  only slightly aspherical, will lead to copious gravitational wave emission \cite{b}. This implies (for the case of bounded sources) that the exterior spacetime should in principle describe such a radiation. However, as is well known, no exact solution, describing gravitational radiation from bounded sources, is available in closed analytical form. The best we have is perhaps the Bondi approach \cite{7, 8} which provides expressions for the metric functions in  terms of inverse power series of the null  coordinate, and whose convergence is only assured very far from the source. In other words, there is not any  explicit exterior metric, to which we could match our interior fluid distribution (in the most general case). In spite of this drawback, we shall be able to provide a formal characterization for the emitted (gravitational) radiation within the source, together with the flow of super--energy associated to the vorticity of the fluid.

An important role in this study is played by a set of scalar functions known as structure scalars. These are obtained from the orthogonal splitting of the Riemann  tensor \cite{9}. They have been shown to be related to fundamental properties of the fluid distribution \cite{10, 11, 12, 13, 14, 15, 16, 17S}. We shall calculate them for our problem. There will be 12 of them, in contrast with the cylindrically symmetric case which is characterized by 8  \cite{14} or the spherically symmetric case, where there is only 5 \cite{9}. We shall relate them to specific physical aspects of the source, and we shall  write down for them the relevant equations. A systematic, though nonexhaustive, study of these equations is carried out. 

Dissipative processes will be treated by means of a causal transport equation derived from the Israel-Stewart theory \cite{18, 19, 20, 21}. This  allows for discussing some interesting  thermodynamic aspects of the problem. Also, its coupling with the generalized ``Euler'' equation, will  illustrate the decreasing of the effective inertial mass density, due to thermal effects, and which may lead to the occurrence of the thermoinertial  bounce. These effects  have already been discussed, in spherically and cylindrically  symmetric systems (see \cite{14, 22, 23, 24, 25, 26, 27, 28, 29, 30} and references therein).

Finally, in the last section, the results will be summarized  and a list of issues deserving further attention will be presented.

\section{The metric and the source: basic definitions and notation}
We shall consider,  axially (and reflection) symmetric sources. For such a system the most general line element may be written in ``Weyl spherical coordinates'' as:

\begin{equation}
ds^2=-A^2 dt^2 + B^2 \left(dr^2
+r^2d\theta^2\right)+C^2d\phi^2+2Gd\theta dt, \label{1b}
\end{equation}
where $A, B, C, G$ are positive functions of $t$, $r$ and $\theta$. We number the coordinates $x^0=t, x^1=r, x^2= \theta, x^3=\phi$.

We shall assume that  our source is filled with an anisotropic and dissipative fluid. We are concerned with either bounded or unbounded configurations. In the former case we should further assume that the fluid is bounded by a timelike surface $\Sigma$, and junction (Darmois) conditions should be imposed there.

The energy momentum tensor may be written in the ``canonical'' form, as 
\begin{equation}
{T}_{\alpha\beta}= (\mu+P) V_\alpha V_\beta+P g _{\alpha \beta} +\Pi_{\alpha \beta}+q_\alpha V_\beta+q_\beta V_\alpha.
\label{6bis}
\end{equation}

The above is the canonical, algebraic decomposition of a second order symmetric tensor with respect to unit timelike vector, which has the standard physical meaning when $T_{\alpha \beta}$ is the energy-momentum tensor describing some energy distribution, and $V^\mu$ the four-velocity assigned by certain observer.

With the above definitions it is clear that $\mu$ is the energy
density (the eigenvalue of $T_{\alpha\beta}$ for eigenvector $V^\alpha$), $q_\alpha$ is the  heat flux, whereas  $P$ is the isotropic pressure, and $\Pi_{\alpha \beta}$ is the anisotropic tensor. We emphasize that we are considering an Eckart frame  where fluid elements are at rest.

Thus, it is immediate to see that 

\begin{equation}
\mu = T_{\alpha \beta} V^\alpha V^\beta, \quad q_\alpha = -\mu V_\alpha - T_{\alpha \beta}V^\beta,\
\label{jc10}
\end{equation} 

\begin{equation}
P = \frac{1}{3} h^{\alpha \beta} T_{\alpha \beta},\quad   \Pi_{\alpha \beta} = h_\alpha^\mu h_\beta^\nu \left(T_{\mu\nu} - P h_{\mu\nu}\right), 
\label{jc11}  
\end{equation}

with $h_{\mu \nu}=g_{\mu\nu}+V_\nu V_\mu$.

Since, we choose the fluid to be comoving in our coordinates, then
\begin{equation}
V^\alpha =(\frac{1}{A},0,0,0); \quad  V_\alpha=(-A,0,\frac{G}{A},0).
\label{m1}
\end{equation}
Next, let us  introduce the unit, spacelike vectors $\bold K, \bold L$, $\bold S$, with components
\begin{equation}
K_\alpha=(0,B,0,0); \quad  L_\alpha=(0,0,\frac{\sqrt{A^2B^2r^2+G^2}}{A},0),
\label{7}
\end{equation}
\begin{equation}
S_\alpha=(0,0,0,C),
\label{3n}
\end{equation}
satisfying  the following relations:
\begin{equation}
V_{\alpha} V^{\alpha}=-K^{\alpha} K_{\alpha}=-L^{\alpha} L_{\alpha}=-S^{\alpha} S_{\alpha}=-1,
\label{4n}
\end{equation}
\begin{equation}
V_{\alpha} K^{\alpha}=V^{\alpha} L_{\alpha}=V^{\alpha} S_{\alpha}=K^{\alpha} L_{\alpha}=K^{\alpha} S_{\alpha}=S^{\alpha} L_{\alpha}=0.
\label{5n}
\end{equation}
The unitary vectors $V^\alpha, L^\alpha, S^\alpha, K^\alpha$ form a canonical  orthonormal tetrad (say  $e^{(a)}_\alpha$), such that  $$e^{(0)}_\alpha=V_\alpha,\quad e^{(1)}_\alpha=K_\alpha,\quad
e^{(2)}_\alpha=L_\alpha,\quad e^{(3)}_\alpha=S_\alpha$$ with $a=0,\,1,\,2,\,3$ (latin indices labeling different vectors of the tetrad). The  dual vector tetrad $e_{(a)}^\alpha$  is easily computed from the condition 

$$ \eta_{(a)(b)}= g_{\alpha\beta} e_{(a)}^\alpha e_{(b)}^\beta$$.

The anisotropic tensor  may be  expressed in the form 
\begin{widetext}
\begin{eqnarray}
\Pi_{\alpha \beta}=\frac{1}{3}(2\Pi_I+\Pi_{II})(K_\alpha K_\beta-\frac{h_{\alpha
\beta}}{3})+\frac{1}{3}(2\Pi _{II}+\Pi_I)(L_\alpha L_\beta-\frac{h_{\alpha
\beta}}{3})+2\Pi _{KL}K_{(\alpha}L_{\beta)} \label{6bb},
\end{eqnarray}
\end{widetext}
with
\begin{eqnarray}
 \Pi _{KL}=K^\alpha L^\beta T_{\alpha \beta} 
, \quad , \label{7P}
\end{eqnarray}

\begin{equation}
\Pi_I=(2K^{\alpha} K^{\beta} -L^{\alpha} L^{\beta}-S^{\alpha} S^{\beta}) T_{\alpha \beta},
\label{2n}
\end{equation}
\begin{equation}
\Pi_{II}=(2L^{\alpha} L^{\beta} -S^{\alpha} S^{\beta}-K^{\alpha} K^{\beta}) T_{\alpha \beta}.
\label{2nbis}
\end{equation}

This specific choice of  these scalars is justified by the fact, that  the relevant equations used  to carry out this study,  become more compact and easier to handle, when expressed in terms of  them.

Finally, observe that from the condition $q^\mu V_\mu=0$, and the fact that due to the symmetry of the problem, Einstein equations imply $T_{03}=0$, it follows
\begin{equation}
q_\mu=q_IK_\mu+q_{II} L_\mu
\label{qn1}
\end{equation}
or, in coordinate components

\begin{equation}
q^\mu=(\frac{q_{II} G}{A \sqrt{A^2B^2r^2+G^2}}, \frac{q_I}{B}, \frac{Aq_{II}}{\sqrt{A^2B^2r^2+G^2}}, 0),\label{q}
\end{equation}
\begin{equation}
 q_\mu=\left(0, B q_I, \frac{\sqrt{A^2B^2r^2+G^2}q_{II}}{A}, 0\right).
\label{qn}
\end{equation}
Of course, all the above quantities depend,  in general, on $t, r, \theta$.
\section{Kinematical variables}
The kinematical variables play an important role in the description of  a self--gravitating fluid. Here, besides the four acceleration, the expansion scalar and the shear tensor, we have a component of vorticity.

Thus we obtain respectively for these variables (see for example \cite{nin})

\begin{eqnarray}
a_\alpha&=&V^\beta V_{\alpha;\beta}=a_I K_\alpha+a_{II}L_\alpha\nonumber\\
&=&\left(0, \frac {A_{,r} }{A },\frac{G}{A^2}\left[-\frac {A_{,t}}{A}+\frac {G_{,t}}{G}\right]+\frac {A_{,\theta}} {A},0\right),
\label{acc}
\end{eqnarray}
\begin{eqnarray}
\Theta&=&V^\alpha_{;\alpha}\nonumber\\
&=&\frac{AB^2}{r^2A^2B^2+G^2}\,\left[r^2\left(2\frac{B_{,t}}{B}+\frac{C_{,t}}{C}\right)\right.\nonumber\\
&&+\left.\frac{G^2}{A^2B^2}\left(\frac{B_{,t}}{B}-\frac{A_{,t}}{A}+\frac{G_{,t}}{G}+\frac{C_{,t}}{C}\right)\right],
\label{theta}
\end{eqnarray}

\begin{equation}
\sigma_{\alpha \beta}= V_{(\alpha;\beta)}+a_{(\alpha}
V_{\beta)}-\frac{1}{3}\Theta h_{\alpha \beta}. \label{acc}
\end{equation}
The non vanishing components of the shear tensor are:
\begin{eqnarray}
\sigma_{11}&=&-\frac{1}{3}\,\frac{1}{r^2A^2B^2+G^2}\,\frac{B^2}{A}\left[r^2A^2B^2\left(-\frac{B_{,t}}{B}+\frac{C_{,t}}{C}\right)\right.\nonumber\\
& & +\left.G^2\left(-2\frac{B_{,t}}{B}-\frac{A_{,t}}{A}+\frac{G_{,t}}{G}+\frac{C_{,t}}{C}\right)\right],\label{sig11}
\end{eqnarray}

\begin{eqnarray}
\sigma_{22}&=&-\frac{1}{3}\,\frac{1}{A^3}\left[r^2A^2B^2\left(-\frac{B_{,t}}{B}+\frac{C_{,t}}{C}\right)\right.\nonumber\\
& & \left. +G^2\left(2\frac{A_{,t}}{A}+\frac{B_{,t}}{B}-2\frac{G_{,t}}{G}+\frac{C_{,t}}{C}\right)\right], \label{sig22}
\end{eqnarray}

\begin{eqnarray}
\sigma_{33}&=&\frac{1}{3}\,\frac{1}{r^2A^2B^2+G^2}\,\frac{C^2}{A}\left[2r^2A^2B^2\left(-\frac{B_{,t}}{B}+\frac{C_{,t}}{C}\right)\right.\nonumber\\
& & \left. +G^2\left(2\frac{C_{,t}}{C}-\frac{B_{,t}}{B}-\frac{G_{,t}}{G}+\frac{A_{,t}}{A}\right)\right].\label{sig33}
\end{eqnarray}
However they are not independent, and therefore the shear tensor may be  defined through two scalar functions, as:

\begin{eqnarray}
\sigma _{\alpha \beta}=\frac{1}{3}(2\sigma _I+\sigma_{II}) (K_\alpha
K_\beta-\frac{1}{3}h_{\alpha \beta})\nonumber \\+\frac{1}{3}(2\sigma _{II}+\sigma_I) (L_\alpha
L_\beta-\frac{1}{3}h_{\alpha \beta}).\label{sigmaT}
\end{eqnarray}

Using (\ref{sig11}), (\ref{sig22}) and (\ref{sig33}) the above scalars may be written in terms of the metric functions and their derivatives as:
\begin{eqnarray}
2\sigma _I+\sigma_{II}&=&\frac{3}{A}\left(\frac{B_{,t}}{B}-\frac{C_{,t}}{C}\right), \label{sigmasI}
\end{eqnarray}
\begin{eqnarray}
2\sigma _{II}+\sigma_I&=&\frac{3}{A^2B^2r^2+G^2}\,\left[AB^2r^2\left(\frac{B_{,t}}{B}-\frac{C_{,t}}{C}\right)\right.\nonumber\\
&
&\left.+\frac{G^2}{A}\left(-\frac{A_{,t}}{A}+\frac{G_{,t}}{G}-\frac{C_{,t}}{C}\right)\right] \label{sigmas},
\end{eqnarray}
 where the comma and the semicolon denote derivatives and covariant derivatives respectively. 
Once again, this specific choice of scalars, is justified by the very conspicuous way, in which they appear in the relevant equations (see the Appendix).

Finally, for the vorticity vector defined as:
\begin{equation}
\omega_\alpha=\frac{1}{2}\,\eta_{\alpha\beta\mu\nu}\,V^{\beta;\mu}\,V^\nu=\frac{1}{2}\,\eta_{\alpha\beta\mu\nu}\,\Omega
^{\beta\mu}\,V^\nu,\label{vomega}
\end{equation}
where $\Omega_{\alpha\beta}=V_{[\alpha;\beta]}+a_{[\alpha}
V_{\beta]}$ and $\eta_{\alpha\beta\mu\nu}$ denote the vorticity tensor and the Levi-Civita tensor respectively; we find a single component different from zero,  producing:

\begin{equation}
\Omega_{\alpha\beta}=\Omega (L_\alpha K_\beta -L_\beta
K_{\alpha}),\label{omegaT}
\end{equation}
and
\begin{equation}
\omega _\alpha =-\Omega S_\alpha.
\end{equation}
with the scalar function $\Omega$ given by
\begin{equation}
\Omega =\frac{G(\frac{G_{,r}}{G}-\frac{2A_{,r}}{A})}{2B\sqrt{A^2B^2r^2+G^2}}.
\label{no}
\end{equation}

Observe that from (\ref{no}) and regularity conditions at the centre, it follows that: $G=0\Leftrightarrow \Omega=0$.

\section{The orthogonal splitting of Riemann  Tensor and structure scalars}
 In this section we shall  introduce a set of scalar functions, known as structure scalars, which are obtained from the orthogonal splitting of the Riemann tensor (see \cite{9, 10, 11, 12, 13, 14} for details). The reason for doing this that we shall express the set of the basic equations deployed in the Appendix, in terms of these scalars.

\noindent Thus, using the Einstein equations, the
Riemann tensor can be decomposed as:

\begin{equation}
R^{\alpha \beta}_{\quad \nu \delta}=R^{\alpha \beta}_{(F)\,\,
\nu\delta}+R^{\alpha \beta}_{(Q)\,\, \nu \delta}+R^{\alpha
\beta}_{(E)\,\, \nu\delta}+R^{\alpha \beta}_{(H)\,\, \nu \delta},
\label{dc1}
\end{equation}
 with
\begin{eqnarray}
R^{\alpha \beta}_{(F)\,\,
\nu\delta}=\frac{16\pi}{3}(\mu+3P)V^{[\alpha}V_{[\nu}
h^{\beta]}_{\delta ]}+\frac{16\pi}{3}\mu h^\alpha_{[\nu}
h^\beta_{\delta ]}, \label{dc2}
\end{eqnarray}
\begin{widetext}
\begin{equation}
R^{\alpha \beta}_{(Q)\,\, \nu \delta}=-16\pi V^{[\alpha
}h^{\beta]}_{[\nu}q_{\delta ]}-16\pi  V_{[\nu}h ^{[\alpha
}_{\delta]}q^{\beta]}-16\pi V ^{[\alpha}V_{[ \nu}\Pi
^{\beta]}_{\delta]}+ 16\pi h^{[\alpha}_{[\nu}\Pi ^{\beta
]}_{\delta]}\label{dc3}
\end{equation}
\end{widetext}
\begin{eqnarray}
 R^{\alpha \beta}_{(E)\,\, \nu\delta}&=&4V^{[\alpha}V_{[\nu} E^{\beta]}_{\delta ]}+
4 h^{[\alpha}_{[\nu}E^{\beta ]}_{\delta ]}, \label{dc4}
\end{eqnarray}
\begin{eqnarray}
 R^{\alpha\beta}_{(H)\,\, \nu \delta}&=&-2 \epsilon ^{\alpha \beta
 \gamma}V_{[\nu}H_{\delta]\gamma}-2\epsilon _{\nu\delta\gamma}V^{[\alpha}H^{\beta
 ]\gamma},\label{dc5}
\end{eqnarray}
 where   $E_{\alpha\beta}$ and $H_{\alpha\beta}$ are the electric
and magnetic parts of the Weyl tensor $C_{\alpha \beta
\gamma\delta}$, defined as usual by
\begin{eqnarray}
E_{\alpha \beta}&=&C_{\alpha\nu\beta\delta}V^\nu V^\delta,\nonumber\\
H_{\alpha\beta}&=&\frac{1}{2}\eta_{\alpha \nu \epsilon
\rho}C^{\quad \epsilon\rho}_{\beta \delta}V^\nu
V^\delta\,,\label{EH}
\end{eqnarray}

\noindent where $\epsilon _{\alpha \beta \rho}=\eta_{\nu
\alpha \beta \rho}V^\nu$ and  subscripts $F, Q, E, H$ have an obvious meaning.

The electric part of the Weyl tensor has only three independent non-vanishing components, whereas only two components define the magnetic part. Thus  we may also write

\begin{widetext}
\begin{equation}
E_{\alpha\beta}=\frac{1}{3}(2\mathcal{E}_I+\mathcal{E}_{II}) (K_\alpha
K_\beta-\frac{1}{3}h_{\alpha \beta}) +\frac{1}{3}(2\mathcal{E}_{II}+\mathcal{E}_{I}) (L_\alpha
L_\beta-\frac{1}{3}h_{\alpha \beta})+\mathcal{E}_{KL} (K_\alpha
L_\beta+K_\beta L_\alpha), \label{E'}
\end{equation}
\end{widetext}
\noindent

and
\begin{equation}
H_{\alpha\beta}=H_1(S_\alpha K_\beta+S_\beta
K_\alpha)+H_2(S_\alpha L_\beta+S_\beta L_\alpha)\label{H'}.
\end{equation}

The orthogonal splitting of the Riemann tensor is  carried out by
means of three tensors $Y_{\alpha\beta}$, $X_{\alpha\beta}$ and
$Z_{\alpha\beta}$ defined as

\begin{equation}
Y_{\alpha \beta}=R_{\alpha \nu \beta \delta}V^\nu V^\delta,
\label{Y}
\end{equation}
\begin{equation}
X_{\alpha \beta}=\frac{1}{2}\eta_{\alpha\nu}^{\quad \epsilon
\rho}R^\star_{\epsilon \rho \beta \delta}V^\nu V^\delta,\label{X}
\end{equation}
and
\begin{equation}
Z_{\alpha\beta}=\frac{1}{2}\epsilon_{\alpha \epsilon \rho}R^{\quad
\epsilon\rho}_{ \delta \beta} V^\delta,\label{Z}
\end{equation}
 where $R^\star _{\alpha \beta \nu
\delta}=\frac{1}{2}\eta_{\epsilon\rho\nu\delta}R_{\alpha
\beta}^{\quad \epsilon \rho}$.

 Using  (\ref{dc1})--(\ref{dc4}) and (\ref{E'}), we obtain:
\begin{eqnarray}
Y_{\alpha \beta}=\frac{1}{3}Y_T h_{\alpha \beta} + \frac{1}{3}(2Y_I+Y_{II}) (K_\alpha
K_\beta-\frac{1}{3}h_{\alpha \beta})\nonumber \\
+\frac{1}{3}(2Y_{II}+Y_{I}) (L_\alpha L_\beta-\frac{1}{3}h_{\alpha \beta})+Y_{KL}
(K_\alpha L_\beta+K_\beta L_\alpha), \label{yf}
\end{eqnarray}
with
\begin{eqnarray}
Y_T=4\pi(\mu+3P), \label{ortc1}\\
Y_I=\mathcal{E}_I-4\pi \Pi_I, \label{ortc2}\\
Y_{II}=\mathcal{E}_{II}-4\pi \Pi_{II}, \label{YY}\\
Y_{KL}=\mathcal{E}_{KL}-4\pi \Pi_{KL}.\label{KL}
\end{eqnarray}

In a similar way the tensor $X_{\alpha \beta}$ can be written as:
\begin{eqnarray}
X_{\alpha \beta}=\frac{1}{3}X_T h_{\alpha \beta} +  \frac{1}{3}(2X_I+X_{II}) (K_\alpha
K_\beta-\frac{1}{3}h_{\alpha \beta})\nonumber \\
+\frac{1}{3}(2X_{II}+X_{I}) (L_\alpha L_\beta-\frac{1}{3}h_{\alpha \beta})+X_{KL}
(K_\alpha L_\beta+K_\beta L_\alpha), \label{xf}
\end{eqnarray}
with
\begin{eqnarray}
X_T=8\pi \mu, \label{ortc1x}\\
X_I=-\mathcal{E}_I-4\pi \Pi_I, \label{ortc2x}\\
X_{II}=-\mathcal{E}_{II}-4\pi \Pi_{II}, \label{YYx}\\
X_{KL}=-\mathcal{E}_{KL}-4\pi \Pi_{KL}.\label{KLx}
\end{eqnarray}

Once again, the  specific choice of all the scalars above has  the purpose of rendering the basic equations in the Appendix in the simpler form.

Finally, from  (\ref{dc1})--(\ref{dc4}), (\ref{EH}) and (\ref{Z})
we obtain
\begin{equation}
Z_{\alpha\beta}=H_{\alpha\beta}+4\pi q^\rho \epsilon_{\alpha\beta
\rho}. \label{Z'}
\end{equation}

or

\begin{equation}
Z_{\alpha\beta}=Z_IK_\beta S_\alpha+Z_{II}K_\alpha S_\beta+Z_{III}L_\alpha S_\beta+Z_{IV}L_\beta S_\alpha \label{Z2}
\end{equation}
where 
\begin{widetext}
\begin{equation}
Z_I=(H_1-4\pi q_{II});\quad Z_{II}=(H_1+4\pi  q_{II}); \quad Z_{III}=(H_2-4\pi q_I); \quad  Z_{IV}=(H_2+4\pi q_I). \label{Z2}
\end{equation}
\end{widetext}

Before ending this section it would be useful to introduce a
relevant quantity defined in terms of  tensors $Y_{\alpha \beta},
X_{\alpha \beta}, Z_{\alpha \beta}$. This is the super--Poynting
vector defined by
\begin{equation}
P_\alpha = \epsilon_{\alpha \beta \gamma}\left(Y^\gamma_\delta
Z^{\beta \delta} - X^\gamma_\delta Z^{\delta\beta}\right),
\label{SPdef}
\end{equation}
which can be written as:
\begin{equation}
 P_\alpha=P_I K_\alpha+P_{II} L_\alpha,\label{SP}
\end{equation}

with
\begin{widetext}
\begin{eqnarray}
P_I &=
&\frac{H_2}{3}(2Y_{II}+Y_I-2X_{II}-X_I)+H_1(Y_{KL}-X_{KL}) + \frac{4\pi q_I}{3}\left[2Y_T+2 X_T-X_I-Y_I\right] \nonumber \\
&-& 4\pi q_{II}(X_{KL} +Y_{KL}),\nonumber
\\
P_{II}&=&\frac{H_1}{3}(2X_{I}+X_{II}-Y_{II}-2Y_I)+H_2(X_{KL}-Y_{KL})-4\pi q_I(Y_{KL}+X_{KL}) \nonumber \\
&+& \frac{4\pi q_{II}}{3}\left[2Y_T+2X_T-X_{II}-Y_{II}\right]. \label{SPP}
\end{eqnarray}
\end{widetext}
Three comments are in order at this point:
\begin{itemize}
\item The super--Poynting vector may be defined in terms of the Riemann tensor (as in (\ref{SPdef})), or in terms of the Weyl tensor \cite{11p, 12p, 14p}. Obviously they coincide in vacuum, but are different within the fluid distribution.
\item  Both components $P_I, P_{II}$ have terms not containing heat dissipative contributions. It is reasonable to associate these with gravitational radiation.
\item Both components of the  super--Poynting vector have contributions of  both components of the heat flux vector.
\end{itemize}
We shall come back to these points, later.
\section{The heat transport equation}
We shall need a transport equation derived from  a causal  dissipative theory ( e.g. the
M\"{u}ller-Israel-Stewart second
order phenomenological theory for dissipative fluids \cite{18, 19, 20, 21}).

Indeed,  the Maxwell-Fourier law for
radiation flux leads to a parabolic equation (diffusion equation)
which predicts propagation of perturbations with infinite speed
(see \cite{63}-\cite{65} and references therein). This simple fact
is at the origin of the pathologies \cite{66} found in the
approaches of Eckart \cite{17} and Landau \cite{67} for
relativistic dissipative processes. To overcome such difficulties,
various relativistic
theories with non-vanishing relaxation times have been proposed in
the past \cite{18, 19, 20, 21, 68, 69}. The important point is that
all these theories provide a heat transport equation which is not
of Maxwell-Fourier type but of Cattaneo type \cite{70}, leading
thereby to a hyperbolic equation for the propagation of thermal
perturbations.

A fundamental parameter   in these theories is the relaxation time $\tau$ of the
corresponding  dissipative process. This positive--definite quantity has a
distinct physical meaning, namely the time taken by the system to return
spontaneously to the steady state (whether of thermodynamic equilibrium or
not) after it has been suddenly removed from it. 
Therefore, when studying transient regimes, i.e., the evolution between two
steady--state situations,  $\tau$ cannot be neglected. In 
fact, leaving aside that parabolic theories are necessarily non--causal,
it is obvious that whenever the time scale of the problem under
consideration becomes of the order of (or smaller) than the relaxation time,
the latter cannot be ignored, since 
neglecting the relaxation time amounts -in this situation- to
disregarding the whole problem under consideration.

Thus, the transport equation for the heat flux reads \cite{19, 20, 21, 64},
\begin{equation}
\tau h^\mu_\nu q^\nu _{;\beta}V^\beta +q^\mu=-\kappa
h^{\mu\nu}(T_{,\nu}+T a_\nu)-\frac{1}{2}\kappa T^2\left
(\frac{\tau V^\alpha}{\kappa T^2}\right )_{;\alpha}q^\mu,\label{qT}
\end{equation}

\noindent where $\tau$, $\kappa$, $T$ denote the relaxation time,
the thermal conductivity and the temperature, respectively.

Contracting (\ref{qT}) with $L_\mu$ we obtain
\begin{widetext}
\begin{eqnarray}
\frac{\tau}{A}\left(q_{II,t}+A q_{I} \Omega\right)+q_{II}=-\frac{\kappa}{A}\left(\frac{G T_{,t}+A^2 T_{,\theta}}{\sqrt{A^2B^2r^2+G^2}}+A T a_{II}\right) -\frac{\kappa T^2q_{II}}{2}(\frac{\tau V^\alpha}{\kappa T^2})_{;\alpha},\label{qT1n}
\end{eqnarray}
\end{widetext}
where  (\ref{no}), has been used

On other hand,  contracting (\ref{qT}) with $K_\mu$, we find

\begin{eqnarray}
\frac{\tau}{A}\left(q_{I,t}-A q_{II} \Omega\right)+q_{I}=-\frac{\kappa}{B}(T_{,r}+BTa_I)\nonumber \\
-\frac{\kappa T^2 q_{I}}{2}(\frac{\tau
V^\alpha}{\kappa T^2})_{;\alpha}. \label{qT2n}
\end{eqnarray}

It is worth noting  that the two equations above are coupled  through the vorticity. We shall discuss futher on this point in the section VII.

\section{Basic equations}
The relevant equations (besides the transport equation shown in the previous section) for describing the evolution of our axially and reflection symmetric dissipative fluid, are obtained applying the 1+3 formalism  \cite{21cil, n2, refe1, n3, n1, refe2, refe3, 22cil, nin} to axial symmetry. Accordingly, they are not new (see for example \cite{n1}), but are exhibited here in the form explicitly adapted to the problem under consideration . The equivalent set of  equations for the spherically symmetric case was obtained in \cite{9}, whereas in \cite{14}, they were obtained for the cylindrically symmetric case.  Obviously, not all of them are independent, however depending on the problem under consideration, it may be more advantageous to use one subset instead of the other, and therefore here we present them all. 
They are presented, with  brief comments about their origins, in Appendix A. The scalar equations obtained by projecting  them on all possible combinations of tetrad vectors $\bold V, \bold K, \bold L, \bold S$, are deployed in the Appendix B.

In what follows we shall extract and discuss part of the information contained in these equations.

\section{Some thermodynamic aspects of the problem}
 The thermodynamics of fluids endowed with vorticity may be quite complicated even in Newtonian theory (e.g. see \cite{dem} for a discussion on this point). However,  even at this level of generality, some interesting conclusions may be drawn from the study of the transport equation  (\ref{qT}) and the generalized ``Euler'' equation (\ref{ec2}).

Thus, as we shall see, the combination of the two above mentioned equations lead to a decreasing of the ``effective'' inertial mass density. This is a known effect, with important implications on the evolution of the object. On the other hand, the fact that both components of  (\ref{qT}) are coupled (through the vorticity), produces a  result which recalls the well known von Zeipel's theorem  \cite{3}. Let us analyze these two issues in some detail.

\subsection{The effective inertial mass density  of the dissipative fluid}
In classical dynamics the inertial mass is defined as the factor of proportionality between the three-force applied to a particle (a fluid element) and the resulting three-acceleration, according to Newton's second law. In relativistic dynamics a similar relation only holds (in general) in the instantaneous rest frame (i.r.f.), since the three-acceleration and the force that causes it are not (in general) parallel, except in the i.r.f.(see for example \cite{Ri}). However, under a variety of circumstances, this factor of proportionality  does not coincide with the mass (density) of the particle (fluid element) in absence of interactions.  In  such cases we refer to this proportionality factor as ``effective inertial mass'' (e.i.m.). Thus for example the e.i.m. of an electron moving under a given force through a crystal, differs from the value corresponding to an electron moving under the same force in free space, and may even become negative (see \cite{Ja, Ki}). 

In our case, combining the equations (\ref{ec2}) and (\ref{qT}) we obtain
\begin{widetext}
\begin{eqnarray}
(\mu+P)\left[1-\frac{\kappa T}{\tau (\mu+P)}\right]a_\alpha=-h^\beta _{\alpha}\Pi ^\mu _{\beta ;\mu}-\nabla _\alpha P-(\sigma _{\alpha \beta}+\Omega _{\alpha \beta})q^\beta\nonumber
\\
+\frac{\kappa}{\tau}\nabla _\alpha T+\left \{\frac{1}{\tau}+\frac{1}{2}D_t \left[ln(\frac{\tau}{\kappa T^2})\right]-\frac{5}{6}\Theta \right\}q_\alpha,
\label{eim}
\end{eqnarray}
\end{widetext}
where $\nabla _\alpha P\equiv h^\beta _\alpha P_{,\beta}$ and $D_t f \equiv f_{,\beta}V^\beta.$

In the above equation we have on the right hand, besides some dissipative terms, terms representing the hydrodynamic ``forces'' acting on any fluid element. On the left hand, it is clear that the factor multiplying the four acceleration vector represents the effective inertial mass density.
Thus, the obtained expression for the e.i.m. density contains a contribution from dissipative variables, which reduces its value with respect to the non-dissipative situation. Such a decreasing of e.i.m. density was  pointed out for the first time in \cite{22}, and since then, it has been shown to appear in a great variety of scenarios (see \cite{15, 23, 26, 28}  and references therein). 

The potential consequences of the above mentioned effect, on the evolution of the self--gravitating object,  should be seriously considered. Indeed, from the equivalence principle it follows that the ``passive'' gravitational mass density should be reduced too, by the same factor  as the e.i.m. density. This in turn might lead, in some critical cases when such diminishing is significative, to a bouncing of the  collapsing object (see \cite{25} for a specific numerical example).

\subsection{Vorticity and heat transport}
As we mentioned earlier, the two components of the transport equation (\ref{qT1n}, \ref{qT2n}), are coupled through the vorticity. This fact entails an interesting thermodynamic consequence.
Indeed, let us assume that at some initial time (say $t=0$) and before it, there is thermodynamic equilibrium  in the  $\theta$ direction, this implies $q_{II}=0$, and  also that the  corresponding Tolman's temperature  \cite{Tol}  is constant, which in turns implies that the term within the round bracket in the first term on the right of (\ref{qT1n}) vanishes. Then it follows at once from (\ref{qT1n}) that:
\begin{equation}
q_{II,t}=-A\Omega q_I,
\label{nvz1}
\end{equation}
implying that the propagation in time of the vanishing of the meridional flow, is subject to the vanishing of the vorticity and/or the vanishing of  heat flow in the $r$- direction.

Inversely, repeating the same argument for (\ref{qT2n}) we obtain at the initial time when we assume thermodynamic equilibrium,

\begin{equation}
q_{I,t}=A\Omega q_{II}.
\label{nvz2}
\end{equation}

Thus, it appears that the vanishing of the radial component of the heat flux vector at some initial time, will propagate in time  if only, the vorticity and/or  the meridional heat flow are different from zero.

In other words, time propagation of the  thermal equilibrium condition, in either direction $r, \theta$, is assured  only in the absence of vorticity. Otherwise, it requires initial thermal  equilibrium in both directions.

This result is a clear reminiscence of the von Zeipel's theorem \cite{3}.
\section{Evolution of the expansion scalar and the shear}
Let us now consider  equations (\ref{esc3})-- (\ref{esc4LL}). In order to elucidate the significance of these  equations, we shall, for simplicity, restrict ourselves to the geodesic fluid ($a_\mu=0$).  The first of these equations, describes  the evolution of  the expansion scalar (Raychaudhuri equation). 

First of all observe that the evolution of the expansion scalar is controlled not only by the scalar $Y_T$  and $\sigma$ as in the cylindrically \cite{14} and the spherically symmetric \cite{9} cases, but also depends on the vorticity vector. It is worth mentioning that, as it is apparent from (\ref{esc51KL})--(\ref{esc6LS}), in the non--geodesic case there is a coupling between $H_1, H_2, \Omega, q_I, q_{II}$, implying that in the general case all these factors also affect the evolution of $\Theta$.

For the shear  we have two equations, (\ref{esc4KK}) and (\ref{esc4LL}) (for the two independent scalars defining  the shear tensor). 

Observe that even in the geodesic case, unlike the cylindrically symmetric case, (\ref{esc4KK}) and (\ref{esc4LL})  are coupled through the $2\sigma^2+\Omega^2$ term.  Thus, assuming that the fluid is initially shear free, the system will deviate from such a condition even  if we keep $Y_I, Y_{II}$ vanishing all along the evolution. In order to keep the fluid shearfree, we need also to keep it, vorticity free. This last condition implies because of (\ref{esc4KL}) that $Y_{KL}$ should also vanish all along the evolution. Thus, the evolution of the shear is now controlled by three structure scalars $Y_I, Y_{II}, Y_{KL}$. In other words all the information about the stability of the shearfree condition is encrypted  in $Y_I, Y_{II}, Y_{KL}$. Once again it should be emphasized that this conclusion is true only for the geodesic case. In the general case, because of (\ref{esc51KL})--(\ref{esc6LS}), we see that the magnetic part of the Weyl tensor and the heat flux vector also affect the stability of the shear--free condition.

\section{Evolution of the vorticity}
Let us now turn to equations (\ref{esc51KL})--(\ref{esc6LS}). If we restrain to the geodesic case, then it seems from (\ref{esc51KL}), that an initially vorticity--free configuration, will remain vorticity--free during the evolution. 
The same situation happens for the shear--free case. 

Indeed, from (\ref{sigmasI}) and (\ref{sigmas}), it follows that the shear--free condition implies 
\begin{equation}
G=ACf(r,\theta),
\label{j1}
\end{equation}
where $f(r,\theta)$ is an arbitrary function of its arguments. Since, neither $A$ nor $C$ can vanish during the evolution, it follows at once from (\ref{j1}) that a shear--free configuration, which is initially vorticity--free, will remain vorticity free during the evolution.

However such  conclusions have to be taken with caution. Indeed, as it follows from (\ref{esc4KL}), the vorticity--free condition implies, in the geodesic case $Y_{KL}=0$. On the other hand as it follows from (\ref{esc6K}) (remember that the metric is non--diagonal and therefore $L^{0}\neq0$), the vorticity--free condition is unstable in the presence of dissipative fluxes, as result of which it appears that the geodesic condition and the shear--free condition, are too restrictive, and  the stability of the vorticity--free condition depends on the above mentioned factors. This fact is in turn, in full agreement with earlier works, where  it was shown that  vorticity generation is sourced by entropy gradients (see \cite{Croco, 82, 81, 83, 75} and references therein).

Finally, observe that if the fluid is shear--free, the vanishing of the vorticity  implies, as it follows from (\ref{esc6KS}) and (\ref{esc6LS}), that the magnetic part of the Weyl tensor  vanishes, too. Also, as it follows from (\ref{esc9S}), the inverse is  true for non--dissipative fluids. This is in full agreement with a result by Glass \cite{glass}, indicating that a necessary and sufficient condition for a shear--free perfect fluid to be irrotational is that the Weyl tensor is purely electric. Thus  we have  extended the Glass result, to  anisotropic fluids.  In the case of dissipative fluids, the vanishing of the magnetic part of the Weyl tensor does not necessarily imply the vanishing of the vorticity.

\section{The density inhomogeneity factors and their evolution}
The density  inhomogeneity factors (in references \cite{9, inh1, 14} they are referred to as inhomogeneity factors), are  specific combinations of physical and geometrical variables (say $\Psi_i$), such that their vanishing is sufficient and necessary condition for the homogeneity of energy density  i.e. $\nabla _\alpha \mu\equiv h^\beta _\alpha \mu_{,\beta}=0$. Of course these latter conditions are necessary but not sufficient for the system to be homogeneous in the broad sense (i.e. a system where spatial gradients of  the Hubble scalar, the pressure, etc, also vanish).

In the spherically symmetric case, in the absence of dissipation, the density  inhomogeneity factor is the scalar associated to the trace--free part of $X_{\alpha \beta}$.  If dissipation is present then additional terms including dissipative flux appear (see \cite{9, inh1} for a detailed discussion).

 In the cylindrically symmetric case,  it was not possible to identify explicitly the density inhomogeneity factors,   nevertheless, it was easy to check that the trace--free part of   $X_{\alpha \beta}$, besides the magnetic part of the Weyl tensor and the dissipative flux determine the energy density inhomogeneity. 

In the static axially symmetric case it was possible to identify the density inhomogeneity factors, they are the structure scalars associated to the trace--free part of  $X_{\alpha \beta}$.

In the present case however, the situation is quite complicated and we were not able to explicitly identify the density inhomogeneity factors. However we can  identify the structure scalars these factors are made of, and their evolution.

Indeed, it follows at once from (\ref{esc8K}) and (\ref{esc8L}) that the vanishing of $X_I, X_{II}, X_{KL}, Z_I, Z_{II}, Z_{III}, Z_{IV}$ implies the homogeneity of energy density (in the sense defined above). On the other hand, the evolution of the above mentioned scalars is determined by (\ref{qT1n}, \ref{qT2n}, \ref{esc7KK}, \ref{esc7KL}, \ref{esc7LL}, \ref{esc7SS}, \ref{esc10SK}, \ref{esc10SL}).

\section{The super--Poynting vector and gravitational radiation}
In the theory of  the super--Poynting vector, a state of gravitational radiation is associated to a  non--vanishing component of the latter (see \cite{11p, 12p, 14p}). This is in agreement with the established link between the super--Poynting vector and the news functions \cite{5p}, in the context of the Bondi--Sachs approach \cite{7, 8}. Furthermore, as it was shown in \cite{5p}, there is always a non-vanishing component of $P^\mu$, on the
plane orthogonal to a unit vector along which there is a non-vanishing component of vorticity (the $\theta-r$- plane).
Inversely, $P^\mu$ vanishes along the $\phi$-direction since there are no motions along this latter direction, because of the reflection symmetry. 

Therefore we can identify three different contributions in (\ref{SPP}). On the one hand we have contributions from the  heat transport process. These are in principle independent of the magnetic part of the Weyl tensor, which explains why they  remain in the spherically symmetric limit. However  the intriguing fact is the appearance of  both components of the four--vector ${\bold q}$ in both components of ${\bold P}$. Observe that this is achieved  through the $X_{KL}+Y_{KL}$ terms in (\ref{SPP}), or using (\ref{KL}, \ref{KLx}), through $\Pi_{KL}$. But we have also seen that  both components of the heat flux vector are coupled through the vorticity, in the transport equation. Thus,  the vorticity  acts as a coupling factor between the two components of the heat flux vector in the transport equation, whereas  $\Pi_{KL}$ couples the two components of the super--Poynting vector, with the two components of the heat flux vector.

On the other hand  we have contributions from the magnetic part of the Weyl tensor. These are of two kinds. On the one hand contributions associated with the propagation of gravitational radiation within the fluid, and on the other, contributions of the flow of super--energy associated with the vorticity on the plane orthogonal to the direction of propagation of the radiation. Both contributions are intertwined, and it appears  to be impossible to disentangle them  through two independent scalars.

It is worth noticing   that the factors multiplying the $H$ terms in  (\ref{SPP}), are $\mathcal{E}_I,  \mathcal{E}_{II}, \mathcal{E}_{KL}$, implying that purely magnetic or purely electric sources, do not produce gravitational radiation. This is consistent with the result obtained in vacuum for the Bondi metric \cite{4p}, stating that purely electric Bondi metrics are static, whereas purely magnetic ones, are just Minkowski.

\section{conclusions and summary of results}
We have carried out a general study on   axially (and reflection)  symmetric relativistic fluids. An important role in this study is played by the structure scalars. 

We have defined the complete set of such scalars corresponding to our problem. 
It turns out that there are twelve  structure scalars ($X_{T, I, II, KL}, Y_{T, I, II, KL}, Z_{I, II, III, IV}$) in contrast with the spherically symmetric  
case where there are only five, and the cylindrically symmetric case where there are only eight.
Besides, two scalars defining the shear tensor ($\sigma_{I,II}$), one scalar defining the vorticity ($\Omega$),  and five  scalars defining the electric and magnetic parts of the Weyl tensor (${\cal E}_{I, II, KL}, H_{1, 2}$) were also introduced.

Next we have identified  and deployed, the set of equations governing   
the structure and evolution of   the system under consideration and brought 
out the role of structure scalars in these equations, in order to exhibit the physical   
relevance of the former.

We have first considered the dynamical equation  (\ref{ec2}) derived from conservation laws, and    coupled it with a transport equation derived from a causal dissipative theory. 
The resulting equation exhibits the decreasing of the effective inertial mass term due to  thermal effects. It is worth noticing that such a decreasing is described by the term in the square bracket on the left hand of side (\ref{eim}), which in turn is produced by the first term on the left and the second term on the right side, of (\ref{qT}) (see \cite{26} for a detailed discussion on this point). But these two terms should enter into any causal and relativistic theory of dissipation. Therefore the effect under consideration is not exclusive to the Israel--Stewart  theory, but must be present in any other reasonable theory of dissipation.

We have also pointed out the coupling of both components of the heat flux vector, through the vorticity. The resulting  situation recalls the picture described by von Zeipel's theorem.

Next, we have  studied the evolution of the expansion scalar, the shear and the vorticity. For simplicity we have considered the geodesic case. Thus we have seen that the evolution of the expansion scalar is controlled by the scalar $Y_T$. However the appearance of the vorticity in the corresponding equation, together with the fact that in the non--geodesic case there is a coupling between $\Omega$ and $Z_{I, II, III, IV}$, leads us to conclude that the latter  scalars also affect the evolution of $\Theta$, if the fluid is not geodesic.

For the shear the situation is similar: in the geodesic case the evolution is controlled by $Y_{I, II, KL}$, however, in the non--geodesic case (by the same reason as in the case of the scalar expansion), the four scalars $Z_{I, II, III, IV}$ are also expected to affect the evolution of the shear.

For the vorticity, it appears that the  geodesic condition may be too stringent. In the general case the evolution of the vorticity depends upon $Y_{KL}$ and $Z_{I, II, III, IV}$.

Next we have considered the density  inhomogeneity factors. Although we were unable to identify these factors explicitly, it was shown that the scalars $X_{I,II, KL}, Z_{I, II, III, IV}$ are the basic constituents of such factors. 

Finally, we analyzed the super--Poynting vector. It contains three types of contributions. On the one hand we have contributions from the dissipative processes associated to the heat flux vector. Next, we have contributions from  gravitational radiation, associated to the magnetic parts of the Weyl tensor. Finally, we have contributions from the  flow of  super--energy, which in turn, acts as the source of the vorticity. 

However, while the pure dissipative contribution is trivially identified, we could not do the same for the other two contributions, since the factors multiplying the $H_1, H_2$ terms in (\ref{SPP}), do not vanish if $\Omega=0$. On the other hand the coupling of both components of the super--Poynting  vector with the two components of the heat flux vector, through $\Pi_{KL}$, appears explicitly  in (\ref{SPP}).

Before ending,   we would like to make some final remarks and to present a partial list of issues, which  remain  unanswered in this manuscript, but  should be addressed in the future. 

\begin{itemize}
\item We have  considered some particular cases, where some variables (e. g. the shear) were considered to vanish. We did so, on the one hand for simplicity, and on the other, in order to bring out the role of some specific variables. However, it should be kept in mind that such kinds of ``suppressions'' may lead to inconsistencies in the set of equations. This is for example the case of ``silent'' universes \cite{s1, s2}, where dust sources have vanishing magnetic Weyl tensor, and lead to a system of 1+3 constraint equations that do not seem to be integrable in general \cite{s3}. In other words for any specific modeling,  the possible occurrence of these types of inconsistencies should be carefully considered.
\item  In the case of specific modeling, another important question arises, namely: what additional information is required to close the system of equations? It is clear that  information about local physical aspects of the source (e.g. equations of state and/or information about energy production) are not included in the set of deployed equations and therefore should be given, in order that metric and matter functions could be solved for in terms of initial data. 

\item From (\ref{SPP}) it follows that either one  of the  ``gravitational'' terms vanish, not only if $H_{1}=0$ or $H_2=0$, but also if, either ${\cal E}_I$,  ${\cal E}_{II}$, or  ${\cal E}_{KL}$ vanish.  What else do these latter conditions imply?
\item Could it be possible to find the exact solution corresponding to nondissipative dust with shear (the analog of the Lemaitre--Tolman--Bondi solution)? Would this solution have a nonvanishing magnetic part of Weyl tensor?
\item Observe that the shearfree condition can be easily integrated from (\ref{sigmasI}) (\ref{sigmas}). Could it be possible to  provide  a comprehensive specific description of shear--free fluids?
\item We have identified the subset of equations which should determine    the density inhomogeneity factors and their evolution, but we were unable to isolate such  factors in the general case. Is this possible?
\\
\item How could one describe the ``cracking'' (splitting) of the configurations as described in \cite{cr1, p3} (and references therein)?
\item As mentioned in the Introduction, we do not have  an exact solution (written down in closed analytical form)  describing gravitational radiation in vacuum, from bounded sources. Furthermore, we do not harbor the hope to find exact analytical solutions, for evolving axially symmetric sources (except perhaps in very restricted situations, e.g. dust). Accordingly, any specific modeling of such a source should be done numerically. 
\item It could be useful to introduce the concept of a mass function, similar to the one existing in the spherically symmetric case. This could be relevant, in particular, in the case of matching the source to a specific exterior.  With respect to this point, it should be mentioned that in  this work we have not considered in detail such a problem, since no specific solution has been presented. However, for any specific model, the correct treatment of such  a matching, would be mandatory.
\end{itemize}
\begin{acknowledgments}
This work was partially
supported by the Spanish Ministry of Science and Innovation (grant
FIS2010-15492) and UFI 11/55 program of the Universidad del Pa\'\i
s Vasco. J.O. acknowledges financial support from the Spanish
Ministry of Science and Innovation (grant FIS2009-07238).
\end{acknowledgments}
\appendix 
\section{The basic equations}
\subsection{Ricci identities} 
From the  Ricci identities for the vector
$V_\alpha$ 
the following set of equations are obtained.

\noindent The time-propagation  equation for the expansion is
$\Theta$
\begin{equation}
\Theta _{;\alpha}V^\alpha +\frac{1}{3}\Theta ^2+2(\sigma ^2-\Omega
^2)-a^\alpha _{;\alpha}+4\pi(\mu+3P)=0\label{ec3}
\end{equation}
\\
 where $2\sigma ^2=\sigma _{\alpha\beta} \sigma
^{\alpha\beta}$.

\noindent The time-propagation equation for the shear is
$\sigma_{\alpha\beta}$
\begin{widetext}
\begin{eqnarray}
h^\mu_{\alpha}h^\nu_{\beta}\sigma_{\mu\nu;\delta}V^\delta+\sigma_\alpha
^\mu \sigma_{\beta \mu}+\frac{2}{3}\Theta \sigma_{\alpha
\beta}-\frac{1}{3}\left ( 2\sigma ^2+\Omega^2-a^\delta
_{;\delta}\right) h_{\alpha \beta}
+\omega _\alpha \omega _\beta-a_\alpha a_\beta-h^\mu
_{(\alpha}h^\nu_{\beta)}a_{\nu;\mu}+E_{\alpha \beta}-4\pi
\Pi_{\alpha \beta}=0,\label{ec4}
\end{eqnarray}
\end{widetext}
and the time-propagation equation for  $\Omega _{\alpha
\beta}$ is

\begin{equation}
 h^\mu _{\alpha}h^\nu _{\beta}\Omega _{\mu\nu;\delta}V^\delta
+\frac{2}{3}\Theta \Omega _{\alpha\beta}
+2\sigma_{\mu[\alpha}\Omega ^\mu
_{\,\,\,\beta]}-h^\mu_{[\alpha}h^\nu
_{\beta]}a_{\mu;\nu}=0.\label{ec51}
\end{equation}

Besides, the following constraint equations follow,
\begin{equation}
h^\beta_\alpha \left (\frac{2}{3}\Theta_{;\beta}-\sigma ^\mu
_{\beta;\mu}+\Omega ^{\,\,\,\mu} _{\beta \,\,\,;\mu}\right )+\left
(\sigma_{\alpha\beta}+\Omega _{\alpha \beta}\right )a^\beta=8\pi
q_\alpha, \label{ec61}
\end{equation}

\begin{equation}
2\omega _{(\alpha }a_{\beta)}+h^\mu _{(\alpha}h_{\beta )\nu}\left
( \sigma_{\mu \delta}+\Omega _{\mu \delta}\right
)_{;\gamma}\eta^{\nu\kappa\gamma\delta}V_\kappa=H_{\alpha
\beta}.\label{ec6}
\end{equation}
\\
\subsection{Conservation laws}
The conservation law $T^\alpha _{\beta;\alpha}=0$ leads to the following equations:
\begin{widetext}
\begin{equation}
\mu _{;\alpha}V^\alpha +(\mu+P)\Theta +\frac{1}{9}(2\sigma _{I}+\sigma _{II})\Pi_I+\frac{1}{9}(2\sigma_{II}+\sigma _I)\Pi_{II}+ q^{\alpha}_{;\alpha} + q^\alpha a_\alpha =0,\label{esc1}
\end{equation}

\begin{equation}
(\mu+P)a_\alpha+h_{\alpha}^\beta\left
(P_{;\beta}+\Pi_{\beta;\mu}^\mu+q_{\beta;\mu}V^\mu\right )+\left(
\frac{4}{3}\Theta h_{\alpha \beta}+\sigma_{\alpha
\beta}+\Omega_{\alpha\beta}\right )q^\beta=0.\label{ec2}
\end{equation}
\end{widetext}
The first of these equations is the ``continuity'' equation, whereas the second one is the ``generalized Euler'' equation.  

\subsection{Differential equations for the Weyl tensor
derived from Bianchi identities}
\noindent From the Bianchi identities
and  Einstein
equations, the following set of equations are obtained:
\begin{widetext}
\begin{eqnarray}
h^\mu_{(\alpha} h^\nu _{\beta)} E_{\mu\nu;\delta}V^\delta+\Theta
E_{\alpha\beta}+h_{\alpha\beta}E_{\mu\nu}\sigma
^{\mu\nu}-3E_{\mu(\alpha}\sigma ^\mu _{\beta )}+h^\mu
_{(\alpha}\eta _{\beta )}^{\,\,\,\, \delta \gamma\kappa}V_\delta
H_{\gamma\mu;\kappa}-E_{\delta
(\alpha}\Omega_{\beta)}^{\,\,\,\delta}\nonumber
\\
-2H^\mu _{(\alpha}\eta_{\beta)\delta \kappa \mu }V^\delta
a^\kappa=-4\pi(\mu+P)\sigma _{\alpha \beta}-\frac{4\pi}{3}\Theta
\Pi_{\alpha \beta}-4\pi h^\mu_{(\alpha} h^\nu_{\beta)}
\Pi_{\mu\nu;\delta}V^\delta-4\pi\sigma_{\mu(\alpha}\Pi_{\beta)}^\mu\nonumber
\\
-4\pi\Omega ^\mu_{\,\,\,(\alpha}\Pi_{\beta )\mu}-8\pi
a_{(\alpha}q_{\beta )}+\frac{4\pi}{3}\left
(\Pi_{\mu\nu}\sigma^{\mu\nu}+a_\mu q^\mu+q^\mu _{;\mu}\right
)h_{\alpha\beta}-4\pi h^\mu_{(\alpha}h_{\beta )}^\nu
q_{\nu;\mu},\label{ec7}
\end{eqnarray}
\end{widetext}
\begin{widetext}
\begin{eqnarray}
h^\mu _\alpha h^{\nu
\beta}E_{\mu\nu;\beta}-\eta_{\alpha}^{\,\,\,\delta \nu
\kappa}V_\delta \sigma ^\gamma _\nu
H_{\kappa\gamma}+3H_{\alpha\beta}\omega ^\beta=\nonumber
\\
\frac{8\pi}{3}h^\beta _\alpha\mu_{;\beta}-4\pi h^\beta_\alpha
h^{\mu\nu}\Pi_{\beta \nu;\mu}-4\pi \left ( \frac{2}{3}\Theta
h_\alpha ^\beta-\sigma ^\beta _\alpha+3\Omega ^{\,\,\,\beta}
_\alpha \right )q_\beta, \label{ec8}
\end{eqnarray}
\end{widetext}
\begin{widetext}
\begin{eqnarray}
\left ( \sigma _{\alpha \delta}E^\delta _{\beta}+3\Omega _{\alpha
\delta}E^\delta _\beta \right)\epsilon _\kappa ^{\,\,\, \alpha
\beta}+a^\nu H_{\nu\kappa}-H^{\nu\delta}_{\,\,\,\, ;\delta}h_{\nu
\kappa} =\nonumber
\\
+4\pi (\mu+P)\Omega _{\alpha\beta}\epsilon _\kappa ^{\,\,\,\alpha
\beta}+4\pi \left [q_{\alpha ;\beta}+\Pi _{\nu\alpha}(\sigma ^\nu
_{\,\,\beta}+\Omega ^\nu _{\,\,\,\beta})\right ]\epsilon _\kappa
^{\,\,\,\alpha \beta},  \label{ec9}
\end{eqnarray}

\begin{eqnarray}
2a_\beta E_{\alpha\kappa}\epsilon_{\gamma}^{\,\,\,\alpha
\beta}-E_{\nu\beta;\delta}h^\nu_{\kappa}\epsilon_{\gamma}^{\,\,\,
\delta \beta}+E^\delta_{\beta;\delta}\epsilon_{\gamma \kappa}^{\quad
\beta}+\frac{2}{3}\Theta H_{\kappa \gamma}+H^\mu_{\nu;\delta}V^\delta
h^\nu _{\kappa}h_{\mu \gamma}\nonumber
\\
-\left (\sigma _{\kappa\delta}+\Omega_{\kappa \delta}\right
)H^\delta _\gamma+\left (\sigma _{\beta \delta}+\Omega _{\beta
\delta}\right )H^\mu _\alpha
\epsilon_{\kappa\,\,\,\mu}^{\,\,\,\delta}\epsilon_{\gamma}^{\,\,\,\alpha\beta}
+\frac{1}{3}\Theta H^\mu _\alpha
\epsilon_{\kappa\,\,\,\mu}^{\,\,\,\delta}\epsilon_{\gamma\,\,\,\delta}^{\,\,\,\alpha}\nonumber
\\
=\frac{4\pi}{3}\mu_{,\beta}\epsilon_{\gamma\kappa}^{\quad \beta}+4\pi
\Pi_{\alpha\nu;\beta}h^\nu_{\kappa}\epsilon_{\gamma}^{\,\,\,\alpha
\beta}+4\pi\left [q_\kappa\Omega_{\alpha \beta}+q_\alpha
(\sigma_{\kappa\beta}+\Omega_{\kappa\beta}+\frac{1}{3}\Theta
h_{\kappa\beta})\right]\epsilon_{\gamma}^{\,\,\,\alpha\beta}.\label{ec10}
\end{eqnarray}
\end{widetext}

 Projecting the equations above, on all possible combinations of the tetrad vectors $\bold{V, K, L, S}$, we find a set of scalar equations, which are deployed in the Appendix B. 
%\appendix 
\section{Summary of scalar equations}
From our basic equations, by projecting on all possible combinations of the tetrad vectors  $\bold{V, K, L, S}$, we find the following scalar equations:

Equation (\ref{ec3})
\begin{equation}
\Theta _{;\alpha}V^\alpha +\frac{1}{3}\Theta ^2+2(\sigma ^2-\Omega
^2)-a^\alpha _{;\alpha}+Y_T=0.\label{esc3}
\end{equation}

Contracting (\ref{ec4}) with $\bold{KK}$, $\bold{KL}$ and $\bold{LL}$ we obtain respectively
\begin{widetext}
\begin{equation}
\sigma _{I,\delta}V^\delta+\frac{1}{3}\sigma
_I ^2+\frac{2}{3}\Theta \sigma_I-(2\sigma ^2+\Omega
^2-a^\delta
_{;\delta})-3(K^\mu K^\nu a_{\nu;\mu}+a_I^2)+Y_I=0,\label{esc4KK}
\end{equation}

\begin{equation}
\frac{1}{3}(\sigma_I-\sigma _{II})\Omega-a_I a_{II}-K^{(\mu}L^{\nu)}a_{\nu;\mu}+Y
_{KL}=0,\label{esc4KL}
\end{equation}
%\end{widetext}
%\begin{widetext}

\begin{equation}
\sigma_{II,\delta}V^\delta+\frac{1}{3}\sigma_{II} ^2+\frac{2}{3}\Theta
\sigma_{II}-(2\sigma ^2+\Omega ^2-a^\delta
_{;\delta})-3(L^\mu L^\nu a_{\nu;\mu}+a_{II}^2)+Y_{II}=0.\label{esc4LL}
\end{equation}
%\end{widetext}

Contracting (\ref{ec51}) with $\bold{KL}$
\begin{equation}
\Omega _{,\delta}V^\delta +\frac{1}{3}(2\Theta+\sigma _I+\sigma _{II})\Omega +K^{[\mu}L^{\nu]}a_{\mu;\nu}=0.\label{esc51KL}
\end{equation}

Contracting (\ref{ec61}) with $\bold{K}$ and $\bold{L}$ we obtain respectively
%\begin{widetext}
\begin{eqnarray}
\frac{2}{3B}\Theta _{,r}-\Omega _{;\mu}L^\mu+\Omega (L_{\beta ;\mu}K^\mu K^\beta-L^\mu _{;\mu})+\frac{1}{3}\sigma _I a_I-\Omega a_{II}-\frac{1}{3}\sigma_{I;\mu}K^\mu\nonumber
\\
-\frac{1}{3}(2\sigma _I+\sigma _{II})(K^\mu _{;\mu}-\frac{a_I}{3})-\frac{1}{3}(2\sigma _{II}+\sigma _I)(L_{\beta ;\mu}L^\mu K^\beta-\frac{a_I}{3})=8\pi
q_I,\label{esc6K}
\end{eqnarray}
%\end{widetext}

%\begin{widetext}
\begin{eqnarray}
\frac{1}{3\sqrt{A^2B^2r^2+G^2}}\left(\frac{2G}{A}
\Theta_{,t}+2A\Theta _{,\theta}\right )+\frac{a_{II} \sigma _{II}}{3}+\Omega _{;\mu}K^\mu+\Omega (K^\mu _{;\mu}+L^\mu K^\beta L_{\beta;\mu})+\Omega a_I-\frac{1}{3}\sigma_{II;\mu}L^\mu \nonumber\\
+\frac{1}{3}(2\sigma _I+\sigma_{II})(L_{\beta;\mu} K^\beta K^\mu+\frac{a_{II}}{3})-\frac{1}{3}(2\sigma _{II}+\sigma _I)(L^\mu _{;\mu}-\frac{a_{II}}{3})=8\pi q_{II}.
\label{esc61L}
\end{eqnarray}
%\end{widetext}

Contracting (\ref{ec6}) with $\bold{KS}$ and  $\bold{LS}$ we obtain respectively:
\begin{equation}
-\Omega a_I-\frac{1}{2}(K^\mu S_\nu+S^\mu K_\nu)(\sigma _{\mu\delta}+\Omega _{\mu\delta})_{;\gamma}\epsilon ^{\nu\gamma\delta}=H_1,\label{esc6KS}
\end{equation}

\begin{equation}
-\Omega a_{II}-\frac{1}{2}(L^\mu S_\nu+S^\mu L_\nu)(\sigma _{\mu\delta}+\Omega _{\mu\delta})_{;\gamma}\epsilon ^{\nu\gamma\delta}=H_2.\label{esc6LS}
\end{equation}

Finally, contracting (\ref{ec7}) with $\bold{KK}$, $\bold{KL}$,  $\bold{LL}$ and $\bold{SS}$ we obtain:
%\begin{widetext}
\begin{eqnarray}
-\frac{1}{3}(X_I-4\pi \mu)_{,\delta}V^\delta+\frac{1}{9}\mathcal{E}_I(3\Theta+\sigma
_{II}-\sigma _I)+\frac{1}{9}(2\sigma _{II}+\sigma _I) \mathcal{E} _{II}\nonumber
\\
-K_\nu \epsilon ^{\nu\gamma\kappa}\left [ H_{1,\kappa}S_\gamma+H_1S_{\gamma ;\kappa}+H_2(S_{\mu;\kappa}L_\gamma K^\mu+L_{\mu;\kappa}S_\gamma K^\mu)\right ]+\Omega X_{KL}\nonumber
\\
=2a_{II}H_1
-\frac{4\pi}{3}(\mu+P+\frac{1}{3}\Pi _I)(\sigma
_I+\Theta)-8\pi
a_I
q_I-\frac{4\pi}{B}(q_I)_{,r} -\frac{4\pi q_{II}A}{\sqrt{A^2B^2r^2+G^2}} \left(\frac{G B_{,t}}{A^2 B} + \frac{B_{,\theta}}{B}\right),\label{esc7KK}
\end{eqnarray}
%\end{widetext}
%\begin{widetext}
\begin{eqnarray}
-X_{KL,\delta}V^\delta+\frac{1}{6}\Omega\left (X _{II}-X_I\right )-\frac{1}{2}X_{KL}(2\Theta-\sigma _I-\sigma
_{II})+a_IH_1-H_2a_{II}\nonumber
\\
-\frac{1}{2}\left [(H_{1,\kappa}S\gamma+H_1(S_{\gamma;\kappa}+S_{\mu;\kappa}K_\gamma K^\mu)+H_2S_\gamma L_{\mu;\kappa}K^\mu)\epsilon ^{\beta \gamma \kappa}L_\beta-(H_1 K^\mu S_\gamma+H_2S^\mu L_\gamma)L_{\mu;\kappa}\epsilon ^{\beta \gamma \kappa}K_\beta\right ]\nonumber
\\
-\frac{1}{2}(H_{2;\kappa}S_\gamma+H_2S_{\gamma ;\kappa})\epsilon ^{\beta \gamma \kappa}K_\beta=\frac{8\pi}{3}\Pi _{KL}(\Theta-\sigma _I-\sigma _{II})-4\pi
a_{II}q_I\nonumber \\
-2\pi(K^\mu L^\nu+K^\nu
L^\mu)q_{\nu;\mu}- 4\pi a_{I} q_{II}, \label{esc7KL}
\end{eqnarray}
%\end{widetext}
%\begin{widetext}
\begin{eqnarray}
\frac{1}{3}(-X _{II}+4\pi\mu
)_{;\delta}V^\delta+\frac{1}{9}\mathcal{E}_{II}
(3\Theta+\sigma _I-\sigma _{II})+\frac{1}{9}(2\sigma
_I+\sigma _{II})\mathcal{E}_{I}
-\Omega X_{KL}\nonumber
\\
-\left [H_{2;\kappa}S_\gamma-H_1(S_\gamma L_{\mu;\kappa}K^\mu+L_{\mu;\kappa}S^\mu K_\gamma)+H_2S_{\gamma;\kappa}\right ]\epsilon ^{\beta\gamma\kappa}L_\beta+2a_{I} H_2\nonumber
\\
=-\frac{4\pi}{3}(\mu+P+\frac{1}{3}\Pi_{II})(\sigma _{II}+\Theta) -8\pi a_{II}q_{II} - 4\pi L^\mu L_\nu q^\nu_{;\mu},\label{esc7LL}
\end{eqnarray}
%\end{widetext}
%\begin{widetext}
\begin{eqnarray}
\frac{1}{3}(X _I+X_{II}+4\pi\mu
)_{;\delta}V^\delta+\frac{1}{3}(X_I+X_{II})
(\Theta+\sigma _I+\sigma _{II})+\frac{1}{9}(2\sigma
_I+\sigma _{II})\mathcal{E}_I
+\frac{1}{9}(2\sigma _{II}+\sigma _I) \mathcal{E}_{II}\nonumber
\\
-(H_{1,\kappa}K_\gamma+H_{2,\kappa}L_\gamma+H_1K_{\gamma;\kappa}+H_2L_{\gamma;\kappa})\epsilon ^{\beta \gamma\kappa}S_\beta+2\left
(H_1  a_{II}-H_2 a_{I}\right)\nonumber
\\
=\frac{4\pi}{3}(\mu+P)(\sigma _I +\sigma
_{II}-\Theta)-\frac{8\pi}{9}(\Theta+2\sigma_I+2\sigma _{II})(\Pi _I+\Pi
_{II})-4\pi q_I \frac{C_{,r}}{BC} - \frac{4\pi q_{II}A}{\sqrt{A^2B^2r^2+G^2}} \left(\frac{G C_{,t}}{A^2 C} + \frac{C_{,\theta}}{C}\right).\label{esc7SS}
\end{eqnarray}
%\end{widetext}
Contraction of (\ref{ec8}) with $\bold{K}$ and $\bold{L}$ produces:
%\begin{widetext}
\begin{eqnarray}
-\frac{1}{3}X_{I,\beta}K^\beta-X_{KL,\beta }L^\beta-\frac{1}{3}(2X_I+X_{II})(K^\beta _{;\beta}-a_\nu K^\nu)-\frac{1}{3}(X_I+2X_{II}) L_{\mu;\beta}L^\beta K^\mu\nonumber
\\
-X_{KL}(L_{\mu;\beta}K^\mu K^\beta+L^\beta _{;\beta}-a_\beta L^\beta)-\frac{1}{3}H_2(\sigma _I+2\sigma _{II})-3\Omega H_1\nonumber \\
=\frac{8\pi}{3}\mu
_{;\beta}K^\beta-\frac{4\pi}{3}q_I(2\Theta-\sigma _I)+ 12 \pi \Omega q_{II},\label{esc8K}
\end{eqnarray}
%\end{widetext}
%\begin{widetext}
\begin{eqnarray}
-\frac{1}{3}X_{II,\beta}L^\beta-X_{KL,\beta }K^\beta-\frac{1}{3}(X_I+2X_{II}) (L^\beta _{;\beta}-a_\beta L^\beta)-\frac{1}{3}(2X_I+X_{II}) K_{\mu;\beta}L^\mu K^\beta \nonumber
\\
-X_{KL}(K_{\mu;\beta}L^\mu L^\beta+K^\beta _{;\beta}-a_\beta K^\beta)+\frac{1}{3}H_1(2\sigma _I+\sigma _{II})-3\Omega
H_2\nonumber \\
=\frac{8\pi}{3}\mu _{;\beta}L^\beta-12\pi\Omega q_I -\frac{4 \pi q_{II}}{3}\left(2 \Theta - \sigma_{II}\right).\label{esc8L}
\end{eqnarray}
%\end{widetext}
Contraction of (\ref{ec9}) with $\bold{S}$ yields:
%\begin{widetext}
\begin{eqnarray}
-\frac{1}{3}X_{KL}(\sigma _{II}-\sigma
_I)+a_IH_1+a_{II} H_2 -H_{1,\delta}K^\delta-H_{2,\delta}L^\delta-H_1(K^\delta _{;\delta}+K^\nu _{;\delta}S^\delta S_\nu)\nonumber
 \\
-H_2(L^\delta _{;\delta}+S^\delta S_\nu L^\nu _{;\delta}) =\left \{8\pi [\mu +P-\frac{1}{3}(\Pi _I+\Pi _{II})]-Y_I-Y_{II}\right \}\Omega
 -\frac{4\pi A (q_IB)_{,\theta}}{B\sqrt{A^2B^2r^2+G^2}}\nonumber \\
 +\frac{4 \pi A}{B \sqrt{A^2B^2r^2+G^2}} \left[\frac{q_{II} \sqrt{(A^2B^2r^2+G^2)}}{A}\right]_{,r},
  \label{esc9S}
 \end{eqnarray}
%\end{widetext}
whereas by contracting  (\ref{ec10}) with $\bold{SK}$ and $\bold{SL}$  we obtain:

%\begin{widetext}
\begin{eqnarray}
-\frac{2}{3} a_{II} \mathcal{E}_I+2a_{I}\mathcal{E}_{KL}-E^\delta _{2;\delta}
L^2-\frac{AY_{I,\theta}}{3\sqrt{A^2B^2r^2+G^2}}+\frac{Y_{KL,r}}{B} \nonumber
\\
-\left [\frac{1}{3}(2Y_I+Y_{II})K_{\beta;\delta}+\frac{1}{3}(2Y_{II}+Y_I)K^\nu L_{\nu;\delta}L_\beta+Y_{KL}(L_{\nu;\delta} K^\nu K_\beta+L_{\beta;\delta})\right ]\epsilon ^{\gamma\delta \beta}S_\gamma\nonumber
\\
+H_{1,\delta}V^\delta +\frac{1}{3}H_1(3\Theta+\sigma _{II}-\sigma _I)+\Omega H_2 =-\frac{4\pi}{3}\mu
_{,\theta}L^2+12\pi \Omega q_I+\frac{4\pi q_{II}}{3}(\sigma _I+\Theta),\label{esc10SK}
\end{eqnarray}
%\end{widetext}

%\begin{widetext}
\begin{eqnarray}
\frac{2a_I}{3}\mathcal{E} _{II}-2a_{II}\mathcal{E}_{KL}+E^{\delta}_{\beta;\delta}K^\beta
+\frac{Y_{II,r}}{3B}-\frac{A Y_{KL,\theta}}{\sqrt{A^2B^2r^2+G^2}}\nonumber
\\
-\left [-\frac{1}{3}(2Y_I+Y_{II})L_{\nu;\delta}K^\nu K_\beta+\frac{1}{3}(2Y_{II}+Y_I)L_{\beta;\delta}+Y_{KL}(K_{\beta;\delta}-K^\nu L_\beta
L_{\nu;\delta})\right ]\epsilon ^{\gamma\delta \beta}S_\gamma\nonumber
\\
+H_{2,\delta}V^\delta +\frac{1}{3}H_2(3\Theta+\sigma _I-\sigma _{II})-\Omega
H_1=\frac{4\pi}{3}\mu_{,\beta}K^\beta-\frac{4\pi q_I}{3}(\sigma _{II}+\Theta)+12\pi \Omega q_{II}.\label{esc10SL}
\end{eqnarray}
\end{widetext}

\end{document}